\documentclass[11pt, prl, prstab, a4paper,showpacs,%
amsmath, amssymb, notitlepage, superscriptaddress]{revtex4}
\usepackage{epsfig}
\usepackage{graphicx}
\usepackage{psfrag}
\usepackage{amssymb}

\newcommand{\refdot}[1]{Ref.~\cite{#1}}
\newcommand{\refsdot}[1]{Refs.~\cite{#1}}
\newcommand{\figdot}[1]{Fig.~\ref{#1}}
\newcommand{\parref}[1]{(\ref{#1})}

\newcommand{\bochum}{Institut f\"ur Theoretische Physik IV,\\%
Fakult\"{a}t f\"ur Physik und Astronomie,\\%
Ruhr-Universit\"{a}t Bochum, D-44780 Bochum, Germany}

\newcommand{\bayreuth}{
Physikalisches Institut, Universit\"{a}t Bayreuth,
D--95440 Bayreuth, Germany
}

\newlength{\figwidth}
\begin{document}

\title{Kinetic electrostatic structures in current-carrying pair plasmas}
\author{A. Luque}
\affiliation{\bochum}
\author{H. Schamel}
\affiliation{\bayreuth}
\author{B. Eliasson}
\author{P.K. Shukla}
\affiliation{\bochum}

\date{\today}
\begin{abstract}
  The existence and properties of phase-space structures in
current-carrying pair plasmas is studied by means of the finite
amplitude expressions of the pseudo-potential method.  Emphasis is
given to double layers, solitary structures and periodic waves.
The results can be applied to electron-positron plasmas as well as to
plasmas containing heavier charged particles with the same mass and
opposite charges, such as fullerene ions and dust particles.  They can
also help to understand numerical simulations that showed the
spontaneous formation of phase-space holes in linearly stable regimes.
\end{abstract}
\pacs{52.35.Fp, 52.35.Sb, 52.27.Ep, 81.05.Tp}
\maketitle
\section{Introduction}
  The investigation of collective phenomena in pair plasmas,
consisting of two species with the same mass is gaining interest among
the plasma physics community.  There are several reasons for that: the
first one is that electron-positron pair plasmas are present in many
astrophysical contexts, such as the early universe, neutron stars and
active galactic nuclei.  On the other hand, recent experiments with
fullerene pair plasmas \cite{Oohara03} have allowed the investigators
to observe collective phenomena in the laboratory, something that was
earlier impeded by the short annihilation times of electron-positron
laboratory plasmas \cite{Surko89, Boehmer95, Liang98}.  But
another explanation for the outstanding interest in pair plasmas is
their relatively simple theoretical description, which makes them a 
particularly good subject of investigations for questions in
fundamental plasma theory.  

  This property of pair plasmas has been emphasized in earlier works
that investigated the nonlinear instability and saturation of
current-carrying plasmas \cite{SchamelLuqueSSR, LuqueSchamelEliassonShukla}.  
In these works, use was made of the equality of time-scales in the dynamics of
both species to accelerate the onset and evolution of instability in a
linearly stable regime.  It was also found that, after a turbulent
transient stage, the system reached a stable state in which phase-space
structures were present coexisting with distribution functions one of
which being characterized by a flat trapped particle distribution.
Kinetic structures in pair plasmas were on
the other hand investigated in Refs.~\cite{SchamelLuquePairPlasma} and 
\cite{EliassonShukla05}, two works on
which the present article relies.  However, in the first, analytical one the
investigations were limited to small 
amplitude waves and a complete symmetry in the parameters describing
the distribution functions of both species whereas in the second, numerical one
only solitary phase-space holes were considered.  Both works did not
consider current-carrying plasmas.  It is thus appropiate to
remove these restrictions in order to arrive at a more general theory
of electrostatic trapping in pair plasmas, which is the purpose of the
present article.

  This paper is organized as follows.  In section \ref{theory} we
provide a theoretical background about the governing equations and the
procedure to obtain finite amplitude equilibrium solutions.  Sections
\ref{DL}, \ref{holes} and \ref{periodic} are dedicated, respectively,
to the study of double layers, phase-space holes and periodic
solutions.  Finally, the results and conclusions are summarized in section 
~\ref{summary}.

\section{Theoretical background}
\label{theory}
  We consider a drifting collisionless pair plasma with equal
temperatures for both particle species.  The dynamics is governed by
the 1D Vlasov-Poisson system
\begin{subequations}
\label{VP}
\begin{equation}
  \left[ \partial_t + v\partial_x \pm \partial_x\Phi(x, t) \partial_v \right]
   f_\mp (x, v, t) = 0,
  \label{vlasov_pair}
\end{equation} 
\begin{equation}
  \partial^2_{xx}\Phi(x, t) = \int dv \, f_- - \int dv \, f_+ \equiv n_- - n_+ ,
  \label{poisson_pair}
\end{equation}
\end{subequations}
where space $x$, time $t$, velocity $v$, the distribution functions
$f_{\pm}$ and electric potential $\Phi$ have been normalized by the Debye 
length $\lambda_D$, the inverse plasma frequency $\omega_p^{-1}$, 
the thermal speed $V_{T}$, $n_0/V_{T}$ and $T/e$,
respectively. Here, $\lambda_D=(n_0 e^2/\epsilon_0T)^{1/2}$, 
$\omega_p=(n_0 e^2/\epsilon_0m)^{1/2}$ and $V_{T}=(T/m)^{1/2}$, where
$n_0$ is the unperturbed particle density of each species, $e$ is 
the magnitude of the electron charge, $m$ is the mass and $T$ is the 
temperature (in Joules) of the two species in the unperturbed state. 
The distribution functions of the homogeneous,
unperturbed state are in the center-of-mass frame 
$f_{0\mp} = (1/\sqrt{2 \pi}) \exp\left\{-(v\mp v_D/2)^2/2\right\}$.

  Electrostatic structures in plasmas have been throughly investigated
by means of the
pseudo-potential method, first introduced in the kinetic regime in
\refdot{Schamel72} and further
developed in Refs.~\onlinecite{Schamel73, Schamel75, Schamel2000} (see also
Refs.~\onlinecite{Schamel86} and \onlinecite{LuqueSchamelReport} for a
review).  In this method, the
distribution functions for equilibria in the wave frame depend on the
constants of motion in a prescribed form, namely
\begin{subequations}
\label{ansatz}
\begin{eqnarray}
f_+(v, \Phi) & = & \frac{N_+}{\sqrt{2\pi}} 
     \nonumber
     \left\{\theta(\epsilon_+) \exp \left[-\frac{1}{2}
      \left(\sigma\sqrt{2 \epsilon_+} + v_+\right)^2\right]
      \right. \\ & & \left.
      + \theta(-\epsilon_+) \exp\left(-\frac{v_+^2}{2}\right)
        \exp\left(-\alpha \epsilon_+\right)\right\},
\end{eqnarray}
\begin{eqnarray}
f_-(v, \Phi) & = & \frac{N_-}{\sqrt{2\pi}} 
     \nonumber
     \left\{\theta(\epsilon_-) \exp \left[-\frac{1}{2}
      \left(\sigma\sqrt{2 \epsilon_-} - v_-\right)^2\right]
      \right. \\ & & \left.
      + \theta(-\epsilon_-) \exp\left(-\frac{v_-^2}{2}\right)
        \exp\left(-\beta \epsilon_-\right)\right\},
\end{eqnarray}
\end{subequations}
where $\theta(z)$ represents the Heaviside step function, $N_\pm$ are
normalization constants,  $v_\mp=v_D/2 \mp v_0$ and $\alpha$ and $\beta$ are 
the trapping parameters of positive and negative ions, respectively. 
The single particle
energies, which are constants of motion, are given by
$\epsilon_\mp := \frac{v^2}{2} - \Phi_\mp$,
where we defined $\Phi_-:=\Phi$ and $\Phi_+:=\Psi - \Phi$
respectively, where $\Psi$ is the maximum value of $\Phi$.
The separatrix in the phase space of
both species is then given by $\epsilon_\mp = 0$, separating free
($\epsilon_\mp > 0$) from trapped ($\epsilon_\mp < 0$) particles. 

  The distribution functions given by \parref{ansatz} can be
integrated in velocity, and yield the particle densities as
functions of the electrostatic potential as
$n_+(\Phi) = N_+ n_0(v_+, \alpha, \Psi - \Phi)$, 
$n_-(\Phi) = N_- n_0(v_-, \beta, \Phi)$, where 
\begin{equation}
	n_0(u, \beta, \Phi) := \exp(-u^2 / 2) 
	  \left[F\left(u^2 / 2, \Phi\right) 
	 	+ T\left(\beta, \Phi\right)\right].
	\label{n}
\end{equation}
The definitions of the special functions $F(v^2/2, \Phi)$ and
$T(\beta, \Phi)$, which represent the contribution of free and trapped
particles, respectively, to the density, are presented in 
\refsdot{Schamel72, Schamel86}.
Note that $n_0(u, \beta, 0) = 1$ for any $u$, $\beta$.
The Poisson
equation is now solved by defining the classical potential
$V(\Phi)$ that satisfies $\Phi''(x) = n_-(\Phi) - n_+(\Phi) =:
-\partial V(\Phi) / \partial \Phi$.  Multiplying both sides by $\Phi'(x)$
and integrating once, we have $\Phi'(x)^2 / 2 + V(\Phi) = 0$, where
the classical potential is
\begin{equation}
	V(\Phi) = N_+ \left[V_0(v_+, \alpha, \Psi) 
		- V_0(v_+, \alpha, \Psi-\Phi)\right]
		- N_- V_0(v_-, \beta, \Phi),
\end{equation}
and we have defined
\begin{equation}
	V_0(u, \beta, \Phi) := \exp\left(-u^2 / 2\right)
		\left[P(\beta, \Phi) - 1 + H(u^2 / 2, 0, \Phi)\right].
\end{equation}
We note that $V(0)=0$.
  The expressions for the special functions $P(\beta, \Phi)$ and
$H(u^2 / 2, 0, \Phi)$ are also found in \refdot{Schamel72, Schamel86}.  
To find
acceptable, physical solutions two conditions have to be imposed upon
$V(\Phi)$: a) $V(\Phi) \le 0$ if $0 \le \Phi \le \Psi$ and b)
$V(\Psi)=0$.  The second condition is usually referred to as the
nonlinear dispersion relation (NDR), as it links the amplitude of the
structure to its phase speed.

\section{Double layers}
\label{DL}
  Now we make use of the expressions presented in the previous section to obtain
equilibrium solutions of the Vlasov-Poisson system.  First, we
look for double layer (DL) solutions.  Double layers are
configurations of phase-space which are associated with a monotonic
step-like potentials \cite{Schamel86, Schamel83}.

  In order to have a DL, the densities of both species must be equal as
$x\to\pm\infty$, which means $\Phi\to 0$ and $\Phi\to\Psi$.  Let us
assume that the densities are unity at $\Phi=0$.  This implies that $N_-=1$ and
$N_+= 1/n_0(v_+, \alpha, \Psi)$.  Under this assumption, the condition
$n_-(\Psi) = n_+(\Psi)$ reduces to
\begin{equation}
	n_0(v_+, \alpha, \Psi) n_0(v_-, \beta, \Psi) = 1.
\end{equation}
  This equation has to be solved simultaneously with the NDR for the unknowns
  $u_0$ and $\Psi$ in order
to obtain valid solutions.  

\subsection{Non-drifting plasma}
\begin{figure}
\includegraphics[width=\figwidth]{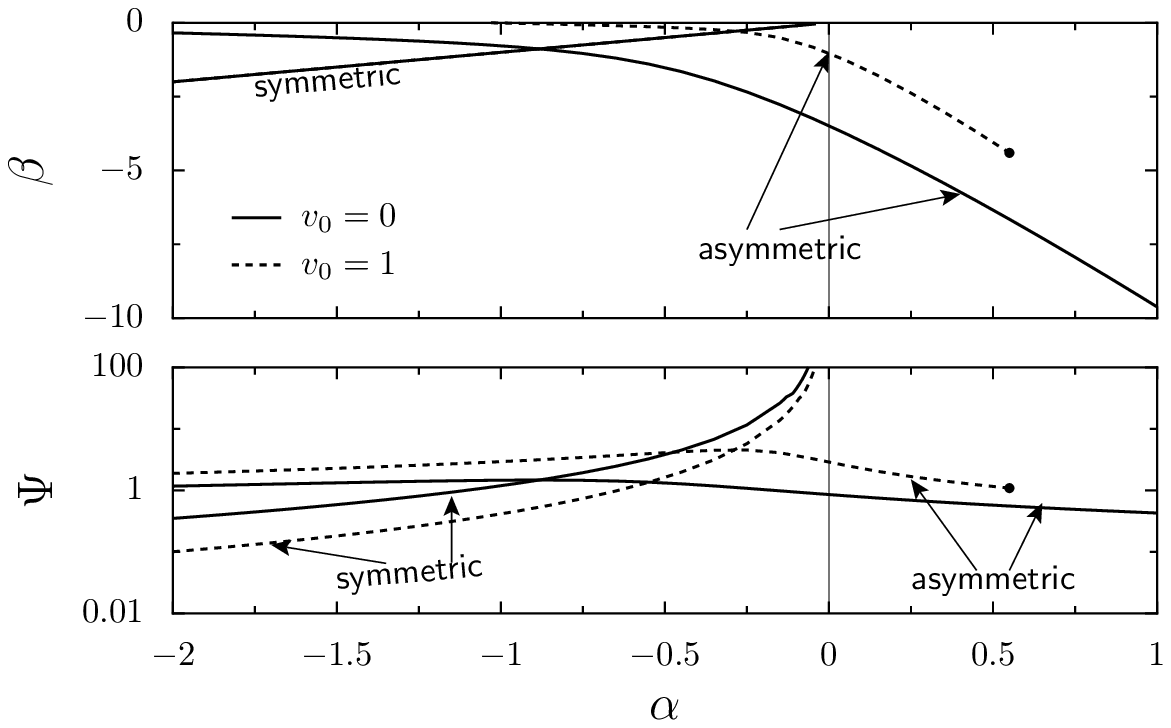}
\caption{Existence curves of double layers (DL) in a non-drifting pair
plasma.  The solid line represents standing DLs while the
dashed one stands for propagating DLs with a phase speed $v_0 = 1$.
Note that for both cases two branches exist, called symmetric
and asymmetric branches.  Note also that there are no asymmetric
solutions in the propagating case if $\alpha > 0.55$.
\label{dlnodrift}}
\end{figure} 

  In \refdot{SchamelLuquePairPlasma}, the existence of DLs
in pair plasmas was discussed.  They were limited, however, to the small
amplitude limit and with the further assumption of complete symmetry between
positive and negative particles ($\alpha=\beta$).  Here we show that,
even if we drop such limitations, we can still find DL
solutions.  Their existence curves are plotted in \figdot{dlnodrift}.
Note that there are two different branches, labelled ``symmetric''
and ``asymmetric''.  

  The symmetric branch represents solutions for which
$\alpha=\beta$. It exists only for $\alpha < 0$ and admits DLs of
arbitrary strength.
In the limit $\Psi\ll 1$ ($|\alpha|=|\beta|
\gg 1$) it corresponds to the solutions of
\refdot{SchamelLuquePairPlasma}, which, for $v_0=0$ satisfy
$(1-\alpha) = 3 \sqrt{\pi}/4\sqrt{\Psi}$.  For finite amplitudes we
observe that the relationship between the defining parameters
is very well approximated by an expression of the form $|\alpha|
\Psi^\gamma = C$, where for standing DLs ($v_0=0$) we found
$\gamma\approx 0.59$, $C\approx 1.08$.  This implies that the DL
becomes the stronger the more flat both trapped particle distributions are.

  The asymmetric branch, on the other side, represents solutions for
which the complete symmetry between the species is broken.  This
branch contains also solutions for which $\alpha \ge 0$.  In the
special case $\alpha=0$, the distribution function of positive
particles is flat in the trapped range while trapped negative
particles form a dip.

  The most relevant physical difference between both branches is that,
while solutions in the symmetric branch do not present a jump in the
densities, this is not true for asymmetric solutions.

\subsection{Current-carrying plasma}
\begin{figure}
\includegraphics[width=\figwidth]{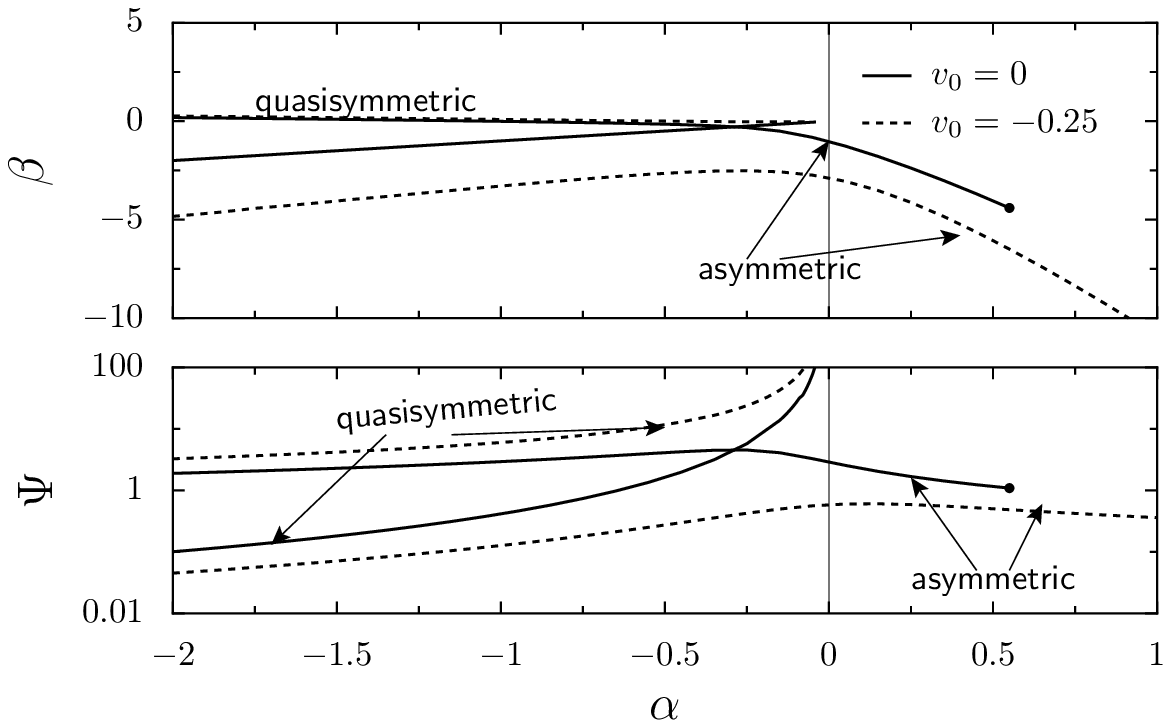}
\caption{Existence curves of double layers (DL) in a current-carrying
pair plasma with $v_D = 2$.
The solid line represents standing DLs while the
dashed one stands for propagating DL with a phase speed $v_0 = -0.25$.
Note that for both cases two branches exist, which are the degenerated
form of the symmetric and asymmetric branches in the case $v_D=0$.  Note 
also that the curves for $v_D=1$, $v_0 = 0$ are equivalent to that of 
$v_D=0$, $v_0=1$, plotted in \figdot{dlnodrift} (see text).
\label{dldrift}}
\end{figure} 

  The picture is changed when we look for double layers in a
current-carrying plasma, a case which was not discussed in
\refdot{SchamelLuquePairPlasma}.  The parameter curves representing
possible solutions for $v_D=2$ are shown in \figdot{dldrift}.

  First of all we note that the
expressions for $n_0(u, \beta, \Phi)$ and $V_0(u, \beta, \Phi)$ do
only depend on $u^2$. Therefore we can make a complete correspondence between
standing structures in a current-carrying plasma ($v_-=v_+=v_D/2$) and
propagating structures in a non-current-carrying plasma ($-v_-=v_+=v_0$) just
by interchanging $v_D/2$ and $\pm v_0$. This explains that the solid
lines in \figdot{dldrift} are the same as the dashed ones in
\figdot{dlnodrift}.  Note however that, although the location of
solutions in parameter space are exactly the same, both represent 
very different kinds of solutions.

  Nevertheless, if we look for propagating structures in
current-carrying plasmas, the symmetry is broken and the
correspondence is no longer valid.  An example of the location in
parameter space of propagating DLs is given by the dashed line in
\figdot{dldrift}.  Note that we still have two different branches but
now, as $v_0 \neq 0$ imposes a further asymmetry, we do not have a
family of solutions with $\alpha=\beta$.  We can still call the two
branches of solutions ``degenerated symmetric branch'' and
``degenerated asymmetric branch''.  The degenerated symmetric branch
exists only for $\alpha < 0$ and requires increasing amplitudes as
$\alpha \to 0$.  Here we find also a power law of the form
$|\alpha|^\gamma \Psi = C$ where for $v_D=2$, $v_0=-0.25$ we obtain
$\gamma \approx 1.015$, $C \approx 6.11$.

\section{Solitary phase-space holes}
\label{holes}
\begin{figure}
\includegraphics[width=\figwidth]{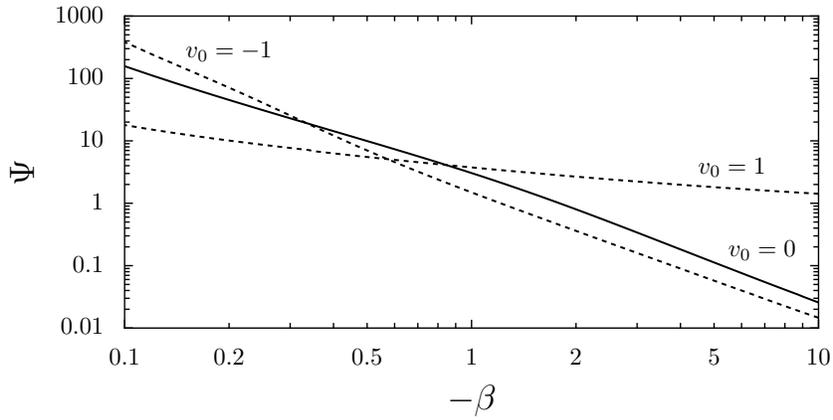}
\caption{The amplitude of the electrostatic potential for holes in
$f_-$ as a function of the trapping parameter $\beta$ for different
phase velocities.  The drift between both species was $v_D=2$ and a flat
trapped range is assumed in $f_+$ ($\alpha = 0$).
\label{ehdrift}}
\end{figure} 

  The existence and properties of solitary holes in non-drifting pair
plasmas was already studied in \refdot{EliassonShukla05}.  Therefore
here we focus on the existence of solitary holes in current-carrying
pair plasmas.  As the symmetry arguments mentioned in the past section
hold for holes as well as for DLs, standing holes in a
current-carrying plasma correspond one-to-one to propagating holes in
a plasma with $v_D=0$.  

  Without loss of generality, we will consider here holes in $f_-$, 
corresponding to bell-like electrostatic potentials, i.e. $V'(0) = 0$, 
$V'(\Psi) > 0$.  We will also consider only the case of a flat trapped
range in $f_+$ ($\alpha = 0$).  There are several reasons to give
special attention to these cases: a) there are large numerical evidences
that such configurations are more stable and therefore potential
attractors of the dynamics.  For example, in
\refdot{EliassonCollidingHoles} the evolution of two colliding holes
was simulated which resulted in a partially flat distribution
function, and b) recent simulations of nonlinear instability
\cite{LuqueSchamelEliassonShukla} and the
subsequent turbulence in current-carrying pair plasmas show that this
turbulence decays towards a stable hole equilibrium in which one of
the species presents a flat distribution in the resonant range.

  In \figdot{ehdrift} we present the curves in the $\beta$, $\Psi$
parameter space where solitary hole solutions exist in a plasma with
$v_D=2$ for different values of the phase velocity $v_0$.  No
solutions are possible with $\beta > 0$.  Note that even for large
amplitudes, the curves can be approximated by a power law of the form
$|\beta|^\gamma\Psi = C$, which is actually exact for small
amplitudes.

\section{Periodic structures}
\label{periodic}
Periodic structures can also appear in pair plasmas.  This is
important because they are excited in experiments such
as \refdot{Oohara03} is usually periodical and also because
most numerical codes impose periodic boundary conditions in space,
giving preference to periodic structures.

In \refdot{SchamelLuquePairPlasma} a dispersion relation was found for
harmonic waves that smoothly joined the limits of
slow acoustic modes $\omega/\sqrt{2} \approx 0.924k$ and
plasma waves $\omega/\sqrt{2} \approx 1$.  To
understand how the finite amplitude of the potential affects these
results, we will use $\alpha = \beta = 1 - v_0^2$, which for
small amplitudes gives raise to harmonic waves ($B = 0$ in
\refdot{SchamelLuquePairPlasma}).

\begin{figure}
\includegraphics[width=\figwidth]{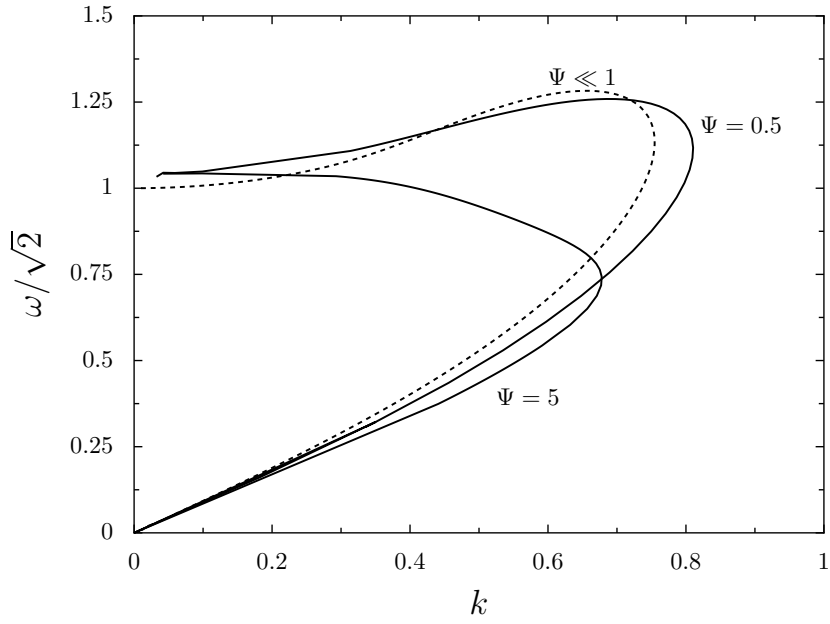}
\caption{Dispersion relation for waves with $\alpha = \beta =
1-v_0^2$ and several amplitudes of the electrostatic potential.  The
dashed curve represents the limit $\Psi \ll 1$, as analytically found
in \refdot{SchamelLuquePairPlasma}.
\label{disprel}}
\end{figure}

\begin{figure}
\includegraphics[width=\figwidth]{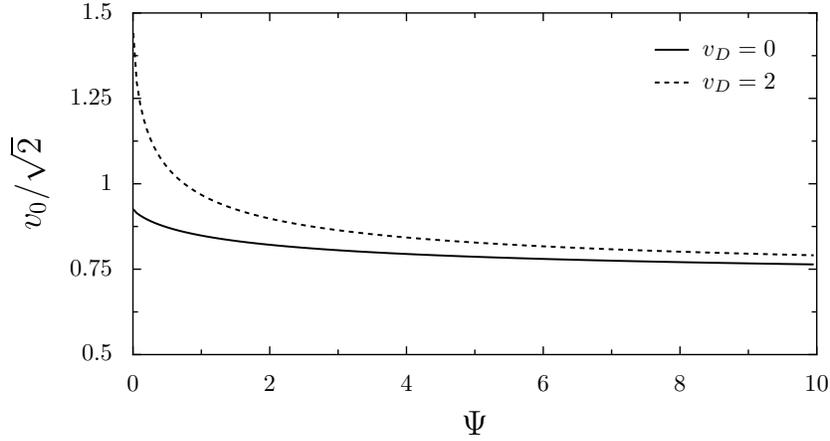}
\caption{Velocity of the slow acoustic branch in the limit of large
wavelengths as a function of the amplitude of the electrostatic potential.
\label{vpsi}}
\end{figure}

Figure~\ref{disprel} shows the dispersion relation of these waves
with different values of the amplitude of the electrostatic
potential.  There we notice that for large amplitude waves
the minimum phase velocity,
corresponding to the slow acoustic mode, is considerably smaller
than the one corresponding to very small amplitudes ($v_0 = 0.924
\sqrt{2}$ when $\Psi \ll 1$).  This slowing effect is shown in \figdot{vpsi},
where the velocity of the slow acoustic mode is plotted as a function
of the amplitude of the electrostatic potential.

\section{Summary and conclusions}
\label{summary}
  In this paper we investigated the existence of electrostatic
structures in pair plasmas with a drift between the species due
e.g. to an external electric field.  This results extend and
complement those of Refs.\cite{SchamelLuquePairPlasma} and
\cite{EliassonShukla05} in several aspects:  we do not restrict
ourselves to the case of perfect symmetry between the species ($\alpha
= \beta$), we do not base the results in the small amplitude limit
($\Psi \ll 1$) and, finally we do not consider only plasmas without a
drift velocity between the species ($v_D = 0$).  By removing those
limitations we have explored a wide range of possible structures. 
We have focused on some remarkable results that may help to interprete future
experimental and numerical data.

  As relevant outcome, we emphasize the existence of
asymmetric double layers, that exhibit a jump in the densities, as well
as the dependence between the velocity of the slow acoustic mode and
the potential amplitude for periodic waves.  This
latter result can easily be compared with experimental data extracted
from a laboratory setup like that of \refdot{Oohara03}, something that we
propose here and that would yield interesting results.

  We would like also to underline that the results presented here are
also connected with the studies about the nonlinear instability and
saturation of pair plasmas, presented elsewhere
\cite{LuqueSchamelEliassonShukla}.  The connection is twofold: a) in
that reference it was shown that nonlinear stability is triggered by
small amplitude phase-space holes and b) the final stable equilibrium
is indeed a structured equilibrium that should be studied with the
tools presented here.

\acknowledgments  This work was supported by the European Commission
(Brussels) through  
contract No. HPRN-CT-2001-00314 for carrying out the task of the RTN
Network ``Turbulent Boundary Layers in Geospace Plasmas'', as well as
by the Deutsche Forschungsgemeinschaft through the
Sonderforschungsbereich 591.


\end{document}